
\hyphenation{cha-rac-te-ris-tic}
\hyphenation{con-fi-gu-ra-tion}
\hyphenation{co-rres-pon-ding}
\hyphenation{TE-NE-RI-FE}
\def\apj{{\it{Ap. J., }}}
\def\apjl{{\it{Ap. J. Letters, }}}
\def\mn{{{Mon. Not. R. astr. Soc., }}}
\def\mn{{\it {MNRAS, }}}

\def\dg{^{\circ }}
\baselineskip=12pt
\def\ref{\hangindent=1cm\hangafter=1}

\def\ref{\noindent\hangindent=1.5cm\hangafter=1}
\def\dg{^{\circ }}
\def\simlt{\lower.5ex\hbox{$\; \buildrel < \over \sim \;$}}
\def\simgt{\lower.5ex\hbox{$\; \buildrel > \over \sim \;$}}
\parskip=0.2 truecm
\parindent=0.5 truecm
\def\cl{\centerline}

\null

\centerline {\bf IMPRINTS OF GALAXY CLUSTERING EVOLUTION ON $\Delta T/T$}
\bigskip
\centerline {Enrique Mart\'\i nez-Gonz\'alez and Jose L. Sanz}
\centerline {Dpto. F\'\i sica Moderna, Universidad de Cantabria and}
\centerline {Instituto de Estudios Avanzados en F\'\i sica Moderna y
Biolog\'\i a Molecular}
\cl {CSIC-Univ. Cantabria, Facultad de Ciencias, Avda. Los
Castros s/n, 39005 Santander, Spain.}
\smallskip
\centerline {Joseph Silk}
\centerline {Departments of Astronomy and Physics, and Center for
Particle Astrophysics,}
\cl {University of California, Berkeley, California 94720}
\vskip 2.truecm

\centerline {\bf ABSTRACT}

The nonlinear evolution of matter clustering induces temperature
anisotropies in the cosmic microwave background. The fluctuations in
temperature are
produced at a relatively low redshift, $z\simlt 10$, and consequently are not
affected by the uncertain reionization proccesses that occurred at earlier
epochs.  The amplitude of the effect depends on the evolution model
assumed for galaxy clustering. For plausible parameter choices, the temperature
fluctuations
are in the range
$10^{-6}\simlt \Delta T/T \simlt 10^{-5}.$
Observed limits on $\Delta T/T $ would be violated if nonlinear clustering on
present cluster scales commenced prior to $z\sim 10.$
\bigskip
\noindent {\it Subject headings:} cosmic background radiation -- cosmology
\vfill\eject
\centerline {\bf I. INTRODUCTION}
\medskip

Most of the studies on the anisotropy expected in the temperature of the
Cosmic Microwave Background (CMB) have been based on linear density
perturbations. Comparatively little attention has been given to additional
effects that can be generated during the nonlinear phase of matter clustering
evolution: the Sunyaev-Zeldovich effect relevant on small angular scales
(Scaramella et al. 1993), bulk motions of gas (Ostriker and Vishniac 1986;
Vishniac 1987; Hu, Scott and Silk 1994) and time-varying gravitational
potentials.
The formalism to calculate the temperature ansotropies generated by
time-varying potentials during the
nonlinear phase has been developed elsewhere (Mart\'\i nez-Gonz\'alez
et al. 1990) for a flat background. Provided that one considers scales much
smaller than the
horizon and non-relativistic matter peculiar velocities, it is possible to
describe the
gravitational field in terms of a single potential $\phi (t; {\bf x})$ that
satisfies the Poisson equation \hfil\break
$${\nabla }^2\phi = {1\over 2}{\rho }_ba^2\triangle (t; {\bf x}),\eqno(1)$$
where $a(t)$ represents the scale factor, ${\bf x}$ are comoving coordinates,
${\rho }_b\propto a^{-3}$ is the background density and $\triangle $ is the
density fluctuation defined by $\rho=\rho_b(1+\triangle).$  We have chosen
units such that $8\pi G = c =1$.

   The temperature anisotropies are given by the expression \hfil\break
$${\Delta T\over T} = {1\over 3}\phi_{LS}+{\bf n\cdot v_{LS}}+2\int_0^{t_{LS}}
dt{\partial \phi \over \partial t}(t; {\bf x}),\eqno(2)$$
where ${\bf n}$ is the unit vector in the direction of the observation and
subscript $LS$ denotes the last scattering surface. In
linear theory, the gravitational potential due to the linear density
fluctuations is static if $\Omega=1,$ so only the two first terms survive
(Sachs and Wolfe
1967) to give the primary anistropies. Unless $\Omega<1,$ the integrated
effect along the path
of the photon  only contributes  as the fluctuations enter the
non-linear regime, to induce secondary anisotropies. The effect of
second--order perturbations on the CMB has been previously studied (Mart\'\i
nez-Gonz\'alez et al. 1992), and were found to contribute to $\Delta
T /T$ at a
level of $\sim 10^{-6}$ for popular models of galaxy formation.
The effect of isolated structures (voids, clusters, great attractors, etc.) has
also been estimated to be at a level of $\Delta T /T\sim 10^{-7}-10^{-6}$
(Rees and Sciama 1968; Thompson and
Vishniac 1987; Kaiser 1982; Nottale 1984; Mart\'\i nez-Gonz\'alez and Sanz
1990). Little attention has been paid to the intermediate regime of weak
clustering.
The effect of clustered matter on the CMB in the very non-linear regime has
been estimated using N-body simulations, for isolated structures (van Kampen
and Mart\'\i nez-Gonz\'alez 1991) and  for the hot dark matter (HDM) scenario
(Anninos et al 1991), where effects of the order of
$10^{-6}$ and $10^{-5}$   are found, respectively (in the latter paper,
they also include Tompson scattering).

\bigskip

\centerline {\bf II. THE NONLINEAR GRAVITATIONAL EFFECT}
\medskip

\centerline {\it a) The effect of a time--varying potential}
\smallskip
     We  consider a flat Friedmann dust model  to
represent our universe from recombination ($z\simeq 10^3$) to the present time
($z=0$). The microwave photons are propagated since the last scattering surface
($z\simeq 10^3$) in any direction and are affected by the gravitational
potential due to the density fluctuations  which are evolving from the linear
regime to the nonlinear one. We are interested in the effect of
nonlinear fluctuations generating anisotropies on the CMB. We calculate this
effect assuming a 2-point density correlation function that can be justified
from observations at the present time and take the time evolution from
numerical simulations.

      We wish to calculate the effect of nonlinear density
fluctuations operating at late times (typically $z\simlt 6$) on scales much
smaller than the horizon. In this case, a non--static potential
$\phi (t,{\bf x})$ suffices to represent the gravitational field of the matter
and its effect on the photons  (Mart\'\i nez-Gonz\'alez et al. 1990), whence
\hfil\break
$${\Delta T\over T} = 2\,\int_e^o dt{\partial \phi\over \partial t}
(t,{\bf x}),\eqno(3)$$
$${\nabla }^2\phi (t,{\bf x}) = 6a^{-1}\triangle (t,{\bf x}),\eqno(4)$$
where $\triangle $ represents the nonlinear density fluctuations. The
integral $(3)$ is extended along the photon path: ${\bf x}(t,{\bf n}) =
\lambda(a)\,
{\bf n}\ \ \ (\lambda (a) = 1-a^{{1\over 2}}$) gives the distance to the photon
from
the observer normalized to the present horizon distance. All units have been
chosen as in the previous section and the scale factor and horizon distance at
the present time are unity, i.e. $a_o = 3\,t_o = 1$.

      The temperature correlation function $C(\alpha )$ is obtained from the
previous equation by averaging over all direction pairs separated by an angle
$\alpha $ \hfil\break
$$C(\alpha ) = 4\,\int_0^{\lambda_e}d\lambda _1\,\int_0^{\lambda _e}
d\lambda _2\,{\partial^2\over \partial \lambda_1\,
\partial \lambda_2}C_\phi (\lambda _1, \lambda _2; x),\eqno(5)$$
$$x({\lambda }_1,{\lambda }_2,\alpha ) \equiv |{\lambda }_1{\bf n}_1-
{\lambda }_2{\bf n}_2| = [{\lambda }_1^2+{\lambda }_2^2-2
{\lambda }_1\,{\lambda }_2\,\cos\alpha ]^{1/2},\eqno(6)$$
where ${\bf n}_1\cdot {\bf n}_2 = cos\alpha $. Thus, $C(\alpha )$ is given in
terms of the correlation of the gravitational potential $C_{\phi }
(\lambda_1, \lambda_2; x)$ at two different cosmic times ${\lambda }_1$ and
${\lambda }_2 $.

\bigskip
\centerline {\it b) Models for the evolution of galaxy clustering }
\smallskip
         We do not have information
 from observations
about the matter
correlations at different cosmic times. Observations of
the galaxy and galaxy cluster distributions are limited to the local universe
with approximately the same cosmic time for all galaxies and clusters.
However, numerical simulations of current models of galaxy formation
(Cole and Efstathiou 1989; Hamilton et al. 1991) imply that there is a simple
time scaling
for the evolution of the matter correlation function. We will assume
the following 2-point correlation function $\xi (z_1, z_2; r)$
$$\eqalign{\xi (z_1, z_2; r)&= [(1+z_1)(1+z_2)]^{-{3+\epsilon \over 2}}
({r_o\over r})^{\gamma },\ \ \ r\leq r_m,\cr
&= 0,\ \ \ r>r_m.\cr}\eqno(7)$$
Here $r$ represents the physical coordinate
and $r_m$ is a
large--scale cut-off beyond which any correlations are assumed to be
vanishingly small.
The previous formula can be justified as follows. The spatial
dependence is obtained from present epoch observations of galaxy clustering,
$\gamma \simeq 1.8$ and $r_o \simeq 5\,h^{-1}\rm Mpc$ (Davis and Peebles 1983;
De Lapparent et al. 1988),
which extend to a scale  $r_m \simeq 15\,h^{-1}\rm Mpc$.
However, positive correlations have recently been found in the APM catalog
(Maddox et al. 1990) up to
angular scales of $\sim 5\dg$ (corresponding to length scales of $\sim 30\,
h^{-1}\rm  Mpc$, given the magnitude limit of the catalog). An equivalent
signal is also apparent  in the counts in cells
of IRAS galaxies of amplitude $\Delta N/N\simeq 0.5 $ in cubes of $30\, h^{-1}
\rm Mpc$ size (Saunders et al. 1991). Additional evidence of clustering up to
scales of $\sim 50\, h^{-1}Mpc$ comes from the distribution of galaxy clusters
(Bahcall and Soneira 1983, Dalton et al. 1992, Peacock and West 1992), although
the amplitude of the cluster autocorrelations is strongly biased.

The evolution of correlations in  numerical simulations
with scale--free spectra as initial conditions, $P(k)\propto k^n, -1\leq n\leq
0$,  can be approximately described by the power law $\xi (z, r) =
(1+z)^{-3}(r_o/r)^{\gamma }$ for scales $r \leq 2\, r_o (1+z)^{-1/\gamma}$
(Cole and Efstathiou 1989)  at late times,
$(z\simlt 6).$ This is equivalent to setting $\epsilon=0$ in (7) and describes
the evolution of  stable clustering where
clusters are already formed and virialized at high $z.$
Less radical evolution of clustering occurs if $\xi(r,z)$ is constant with
time in comoving coordinates. In this case, the clustering developed early and
froze out, as would be appropriate for a biased galaxy formation model. In
this case,.
$\epsilon=\gamma -3.$ It is also possible to have  $\epsilon > 0$ in models
where clusters form more recently and are still evolving
at present. Some N-body simulations (Melott 1992) predict $0\leq\epsilon \leq
3$
for scenarios with spectral indices in the range $-3 \leq n \leq 1$.

Wolfe (1993) has recently found that the clustering of damped Ly$\alpha$
absortion systems ($z\approx 2-3$) in quasars is best fitted with an evolution
exponent $\epsilon=-1.2\, \ (\gamma=1.8)$.
In what follows, we will not choose any specific value of
$\epsilon,$  but  we will consider values within a  broad
interval ($-1.2\le \epsilon\le 2$) that covers the range of interest.
Roughly speaking, values of
$\gamma-3 \le \epsilon \le 0$ will represent different evolutionary models
where clusters form very early, whereas $\epsilon>0$ describes late and
ongoing clustering, as would be appropriate, for example, if $\Omega<<1$ and
$\Omega=1$, respectively.

The above
expression for $\xi $ can be rewritten in terms of $\lambda $ and comoving
distance $x$ as \hfil\break
$$\eqalignno{\xi (\lambda_1, \lambda_2; x)&= [(1-\lambda _1)(1-\lambda _2)]^{3-
                                 \gamma+\epsilon }({r_o\over x})^{\gamma },\ \
\
                                 x\leq x_m,\cr
                                        &= 0\ \ \ x> x_m.&(8)\cr}$$
We will test this type of 2-point correlation function in terms of the
anisotropies generated in the CMB. The {\it ad hoc} cut-off $x_m$ is
time--dependent and represents the maximum scale where the  nonlinear evolution
model is appropriate: $\xi (\lambda_1, \lambda_2 ; x) = \xi_c;\ \xi_c\simgt 0.1
$. We write
\hfil\break
$$x_m = r_o\,\xi_c^{-{1\over \gamma }}[(1-\lambda _1)(1-\lambda _2)]
             ^{{3-\gamma +\epsilon\over \gamma }}.\eqno(9)$$
Thus we have three phenomenological parameters to describe clustering
evolution: the initial redshift for the start of
nonlinear evolution $z_m$ ($\lambda_m =
1-(1+z_e)^{-1/2}$), the evolution parameter $\epsilon $ and the correlation
cut--off $x_m$ at present, $x_{mo}\equiv r_m=r_o\, \xi_c^{-{1\over \gamma }}$.
\bigskip

\centerline {\it c) $\Delta T/ T$ for models of galaxy clustering}
\smallskip
 From the Poisson equation (4) one obtains \hfil\break
$${\nabla }_{\bf x_1}^2{\nabla }_{\bf x_2}^2\,C_{\phi }
(\lambda_1,\lambda_2; x)= 36\,[(1-{\lambda }_1)(1-{\lambda }_2)]^{-2}
\xi ({\lambda }_1,{\lambda _2}; x)\ ,\eqno(10)$$

Assuming standard boundary conditions: i) $|C_{\phi
}|<\infty $, ii) $C_{\phi }(\infty ) = 0$, iii) $C^{\prime }_{\phi }(0) = 0 $
and iv) continuity up to the second derivative of $C_{\phi }$,
the correlation of the gravitational potential, associated to the
correlation of matter given by equation (8), is unique and defined by the
expression \hfil\break
$$\eqalignno{C_{\phi }&= f\,g\,x_o^{\gamma }x_m^{4-\gamma }\big\{ 3[(4-\gamma )
		    (3-\gamma )
                    -2]-(5-\gamma )(4-\gamma )({x\over x_m})^2+6({x\over x_m})^
                    {4-\gamma }\big\},\ \ \ \ x\leq x_m,\cr
                  &= f\,g\,x_o^{\gamma }x_m^{4-\gamma }{2(4-\gamma )
                    (2-\gamma )({x_m\over x})},\ \ \ \ x>x_m,&(11)\cr}$$
where
$$f = [(1-\lambda_1)(1-\lambda_2)]^{1-\gamma -\epsilon },\
g = -{1\over 5-\gamma }+{3\over 4-\gamma }-{3\over 3-\gamma }+
            {1\over 2-\gamma }\, .\eqno(12)$$

In the next section we illustrate these results for the potential correlation
, $C_\phi(x)$, and
the temperature correlation, $C(\alpha)$ (as given by equations 5 and 11), for
several cases of galaxy clustering evolution.
\bigskip

\centerline {\bf III. RESULTS}
\medskip

	The behaviour of the potential correlation function $C_\phi (x)$ is shown in
Fig. 1 as a function of the comoving distance $x$  for several
values of the evolution parameter $\epsilon$ and intial redshift $z_m$. The
correlations grow with $z$ for comoving clustering ($\epsilon =-1.2$) whereas
they decrease for stable clustering ($\epsilon =0$) and for more rapid
evolution ($\epsilon > 0$). This behaviour of $C_\phi$ implies that the
nonlinear effect on $\Delta T/T$  depends on the initial redshift at which
the nonlinear evolution of matter initially occurs for comoving clustering.
For stable and faster clustering evolution, the effect will be produced at
low
redshift $z\simlt 3$ even if we extrapolate the power law evolution (eq. 8)
to arbitrarily high $z$.

	Fig. 2 shows the temperature correlation function generated by the
nonlinear evolution of matter given by eq. (8) for three cases of the
evolution parameter $\epsilon=-1.2,\, 0,\, 2$. The correlation scale is of the
order of a degree and is larger for higher $\epsilon$. In the case of comoving
clustering (fig. 2a), most of the effect is produced close to the initial
redshift $z\simlt 10.$  However for stable clustering the temperature
correlations are generated in the range
$3\simlt z\simlt 0.5,$ (fig. 2b) and for $\epsilon=2$ in the range $0.5\simlt
z \simlt 0.1$ (fig. 2c).

We also find that $C(\alpha)$ is very sensitive to the matter correlation
cut-off
$\xi_c$. This is shown in fig. 2d,e,f for $\xi_c=0.287 (r_m=10\,h^{-1}\rm
Mpc),0.2,
0.1$.

    The main results that we have obtained for the different models of galaxy
clustering evolution can be summarized as follows:

\item{(a)} The temperature correlations $C(\alpha )^{{1\over 2}} \simgt
10^{-6}$ for values of the evolution parameter $-1.2 \leq \epsilon
\leq 3$, initial redshift $z_m\geq 3$ and cut-off $r_m\geq 10\, h^{-1}Mpc$
at present (i.e. $\xi_c \leq 0.287$).

\item{(b)} The effect on $\Delta T/ T$ reaches a maximum for comoving
clustering
($\epsilon =-1.2$), decreases for stable clustering ($\epsilon =0$) and slowly
increases for $\epsilon > 0$.

\item{(c)} More than $80\% $ of the nonlinear effect is produced in the
interval $z \simlt 3$ for $\epsilon \simgt 0$ whereas for $\epsilon = -1.2$
the effect occurs near  the initial redshift when comoving nonlinear
evolution commences.

\item{(d)} Considering present experimental limits on  degree scales,
the evolution of the correlations for comoving clustering as given by eq. (8)
(with $r_m=10\, h^{-1}\rm Mpc$)
must have commenced at $z\simlt 10$, otherwise the present limits on $\Delta
T/T$ of $\approx 10^{-5}$ would be violated.
Extrapolating the observed galaxy correlation power
law up to $r_m=19.4\, h^{-1}\rm Mpc$ ($\xi_c=0.1$) those limits imply that
clusters must have formed later than $z\simeq 4$ for the case of comoving
evolution.
\bigskip

\centerline {\bf IV. CONCLUSIONS}

    We have shown that galaxy clustering evolution typically generates CMB
temperature fluctuations in the range $10^{-6}\simlt \Delta T/T \simlt
10^{-5}$.
For some specific models, an
observable effect on degree scales can be obtained. This anisotropy is
generated at redshifts below $\sim 10$ and therefore is not erased by
reionization processes that could take place in the universe at high $z$.

    Observational limits on the anisotropies of the CMB on degree scales
constrain models for the evolution of galaxy clustering and therefore of
cluster formation. The maximum anisotropy is generated in the case of
comoving clustering where the limits on $\Delta T/T$ are violated if
we extrapolate the observed galaxy correlation power law up to $19.4\, h^{-1}
Mpc$ (corresponding to a correlation $\xi_c=0.1$) with $z_m=6$ or allowing
$z_m\simgt 10$ and keeping the cut-off at $10\, h^{-1} Mpc$ ($\xi_c=0.287$).
The stable clustering model
($\epsilon =0$) or models with a faster evolution ($\epsilon >0$) contribute
to $\Delta T/T$ at a level of the order of $10^{-6}$
whereas models that form structure at high-z are
constrained by the CMB observations.

It would be very interesting to calculate the nonlinear gravitational effect
using more realistic models for the evolution of galaxy clustering that
smoothly
interpolate between the linear and nonlinear regimes. Such a calculation would
clarify the nonlinear versus linear contributions. Indeed, from the
numerical simulations done by Anninos et al. (1991) for a HDM model, it
is seen that the nonlinear gravitational effect can be of the same order
as the linear Sachs-Wolfe effect on degree angular scales.
\bigskip
\bigskip
\bigskip
    E. M.-G. and J. L. S. wish to acknowledge the hospitality of the Berkeley
Astronomy Department and the CfPA where part of this work was made. They
thank the Commission of the European Communities, Human
Capital and Mobility Program of the EC, contract number ERBCHRXCT920079, and
the Spanish DGICYT, project number PB92-0434-C02-02 for
partial financial support. This work was supported at Berkeley in part by
grants from DOE and NSF. We thank W. Hu for a critical reading of the
manuscript.
\bigskip
\centerline {\bf REFERENCES}

\ref Anninos, P., Matzner, R.A., Tuluie, R. \& Centrella, J. 1991, \apj
{\bf 382}, 71

\ref Bahcall, N.A. \& Soneira, R.M. 1983, \apj {\bf 270}, 20


\ref Cole, S. \& Efstathiou, G. 1989, \mn {\bf 239}, 195

\ref Dalton, G.B., Efstathiou, G., Maddox, S.J. \& Sutherland, W.J. 1992,
\apjl {\bf 390}, L1

\ref De Lapparent, V., Geller, M. J. and Huchra, J. P. 1988, \apj {\bf 332}, 44


\ref Hamilton, A.J.S., Kumar, P., Lu, E. \& Matthews, A. 1991, \apjl {\bf 374},
L1

\ref Hu, W., Scott, D. \& Silk J. 1994, \apj (in press)

\ref Kaiser, N. 1982, \mn {\bf 198}, 1033

\ref van Kampen, E. \& Mart\'\i nez-Gonz\'alez, E. 1991, in Proceedings of
Second Rencontres de Blois "Physical Cosmology", ed. A. Blanchard, L.
Celnikier, M. Lachieze-Rey \& J. Tran Thanh Van (Editions Frontieres), p 582

\ref Mart\'\i nez-Gonz\'alez. E., Sanz, J. L. 1990, \mn {\bf 247}, 473

\ref Mart\'\i nez-Gonz\'alez, E., Sanz, J. L. \& Silk J. 1990, \apjl {\bf 355},
L5

\ref -----------------------. 1992, {\it Phys. Rev. D}, {\bf 46}, 4193

\ref Maddox, S.J., Efstathiou, G., Sutherland, W.J. \& Loveday, J. 1990, \mn
{\bf 242}, L43


\ref Melott, A.L. 1992, \apjl {\bf 393}, L45

\ref Nottale, L. 1984, \mn {\bf 206}, 713

\ref Ostriker, J.P. \& Vishniac, E.T. 1986, \apjl {\bf 306}, L51

\ref Peacock, J.A. \& West, M.J. 1992, \mn {\bf 259}, 494

\ref Sachs, R.K. \& Wolfe, A.N. 1967, \apj {\bf 147}, 73

\ref Scaramella, R., Cen, R. \& Ostriker, J.P. 1993, \apj {\bf 416}, 399


\ref Thompson, K.L. \& Vishniac, E.T. 1987, \apj {\bf 313}, 517

\ref Vishniac, E.T. 1987, \apj {\bf 322}, 597

\ref Wolfe, A.M. 1993, {\it Annals of the New York Academy of Sciencies},
{\bf 688}, 281

\vfill\eject
\centerline {FIGURE CAPTIONS}
\bigskip
\noindent {\bf Figure 1.} Potential correlation function $C_\phi (x)$ as a
function of the comoving distance $x$ for three values of the evolution
parameter: a) $\epsilon =-1.2$ (comoving clustering), b) $\epsilon= 0$
(stable clustering) and c) $\epsilon =2$. Different lines correspond to
the following redshifts for the start of nonlinear evolution: $z_m=0$ (solid),
$0.5$ (dotted), $1$ (short-dashed), $3$ (long-dashed) and  $6$ (dash-dotted).
The correlation cut-off is assumed to be $\xi_c=0.287$.
\medskip
\noindent {\bf Figure 2.} Temperature correlation function $C(\alpha)$ as
a function of the angular distance $\alpha$. a) $\epsilon=-1.2$ and
$z_m=10, 6, 3, 1$ (solid, dotted, short-dashed, long-dashed), b) $\epsilon=0$
and $z_m= 6, 3, 1$
(solid, dotted, short-dashed) and c) $\epsilon=2$ and $z_m= 6, 0.5, 0.1$
(solid, dotted, short-dashed).
The correlation cut-off is assumed to be $\xi_c=0.287$ for cases a,b,c.
d) $\epsilon=-1.2$, e) $\epsilon=0$, f) $\epsilon=2$, and $\xi_c=0.287, 0.2,
0.1 (r_m=10,12.2,19.4\, h^{-1}Mpc)$
for solid, dotted and dashed line
respectively. The starting redshift $z_m=6$ is the same for cases d,e,f.

\end

\magnification=\magstep1
\parskip=0.2truecm
\parindent=.5truecm
\raggedbottom

\newcount\fcount \fcount=0

\def\ref#1{\global\advance\fcount by 1 \global\xdef#1{\relax\the\fcount}}
\def\pp{\parshape 2 0truecm 16.5truecm .5truecm 14.5truecm}

\def\book #1;#2;#3{\par\pp #1, {\it #2}, #3}
\def\rep #1;#2;#3{\par\pp #1, #2, #3}

\def\simlt{\lower.5ex\hbox{$\; \buildrel < \over \sim \;$}}
\def\simgt{\lower.5ex\hbox{$\; \buildrel > \over \sim \;$}}

\def\cl{\centerline}

\def\lsim{\hbox{\raise 1 truemm \rlap{$<$}\lower 1truemm \hbox{$\sim$}\ }}
\def\lsim{\hbox{\raise 1 truemm \rlap{$>$}\lower 1truemm \hbox{$\sim$}\ }}
